# Nonlinear Parity-Time (PT) symmetric closed-form optical quadrimer waveguides: Attractor perspective

Samit Kumar Gupta, Jyoti P. Deka and Amarendra K. Sarma*

Department of Physics, Indian Institute of Technology Guwahati, Guwahati-781 039, Assam, India
*Electronic address: aksarma@iitg.ernet.in

**Abstract**. We report a study on a closed-form nonlinear parity-time symmetric optical quadrimer waveguides system with a specific coupling scheme. The system yields power saturation behavior in the modes, which may be attributed to the inherent attractor in the system. A detailed analysis has been provided to confirm the attractor aspect of the system. This work also addresses a crucial issue regarding choice of initial conditions while carrying out numerical simulation for such systems.

## 1 Introduction

Recently, study of parity-time symmetry in optics has become a topic of considerable significance in both fundamental and applied research [2-18]. It was the pioneering work of Bender and Boettcher, where they showed that a wide class of non-Hermitian Hamiltonians could exhibit entirely real spectra as long as they respect the conditions of parity and time (PT) symmetry [1]. For example, the well-known Hamiltonian in quantum mechanics, describing the so-called Schrodinger equation is PT-symmetric provided $V(x) = V^*(-x) = V_R(x) + i\epsilon V_I(x)$ and shows real eigen-spectra below a certain threshold, say $\epsilon < \epsilon_{th}$. If $\epsilon > \epsilon_{th}$, the so-called PT-symmetry breaking threshold, the PT symmetry will spontaneously break down and the spectrum will cease to be entirely real. Taking a hint from the fact that, in optics, the paraxial equation of diffraction is mathematically isomorphic to the Schrödinger equation in quantum mechanics [17], it is possible to study PT-symmetry in optics if one can appropriately tailor the refractive index profile, $n(x)$, of the system such that $n(x) = n^*(-x)$. This means that the real part the refractive index distribution must be an even function of position whereas the gain/loss distribution must be odd. Such PT-systems are characterized by simultaneous presence of balanced gain and loss into the systems and they can give rise to a phase transition (spontaneous PT-symmetry breaking) from an unbroken PT-phase to a broken PT-phase, once the gain/loss parameter exceeds a critical value. Above this threshold value, in spite of the fact that the necessary condition of parity-time symmetry, i.e. $[H, PT] = 0$ is still satisfied, the eigen-values of the system become complex or imaginary. Many such optical systems have been artificially designed. Parity-time symmetry has been realized experimentally in various optical systems in the last few years [19-25]. It is worthwhile to mention that the possibility of exploring the idea of parity-time symmetry in optics was first proposed by A. Ruschhaupt, F. Delgado and J. G. Muga in 2005[18].

In the recent past, we have seen enormous investigations being carried out in linear and nonlinear optical structures themed upon PT-symmetry both on theoretical and experimental grounds. Instead of having passive optical structures, the idea of having a system of exactly balanced gain and loss has triggered revelation of many novel features not found in normal passive systems, such as: unidirectional invisibility [7, 26], non-reciprocity and power oscillation [27], coherent perfect absorbers [28, 29], broad-area single mode lasers [10] etc. In nonlinear optics, parity-time (PT)-symmetry has been explored in diverse contexts, such as: stable dark solitons in dual-core waveguides [30], dynamics of a chain of interacting PT-invariant nonlinear dimers [31], field propagation in linear and nonlinear stochastic PT-symmetric coupler [32], ghost states in PT-symmetric birefringent optical coupler [33], stability of solitons in PT-symmetric coupler [34],PT-symmetry breaking in a necklace of coupled optical waveguides [35], Bragg solitons in nonlinear PT-symmetric periodic potential [36], continuous and discrete Schrödinger systems with parity-time symmetric nonlinearities [37], instabilities, solitons and rogue waves in PT-symmetric coupled nonlinear waveguides[38], bright and dark solitary waves and existence of optical rogue waves [39], modulation instability in nonlinear complex parity-time (PT)-symmetric periodic structures [40] and so on. On the other side, the PT-symmetric oligomers such as dimers, trimers and quadrimers are still drawing considerable amount of research interests owing to possible realization of novel and complex PT-symmetric structures and lattices, using these as building blocks. In particular, PT symmetric quadrimer system has been studied by many groups in various contexts [41-51]. In this work, we have considered a closed-form parity-time symmetric quadrimer waveguides structure with a specific coupling scheme. This specific coupling yields unique linear eigen-spectrum subject to choice of various coupling parameters. On the other side, in nonlinear regime, a detailed nonlinear analysis has been conducted to characterize the power saturation behaviors in the different sites. Upon thorough inspection, we confirm that the saturation points correspond to the attractors of the system. It is noteworthy that the power saturation behavior in oligomers system has been reported in completely different settings, for example, dimer and trimer systems with linear and nonlinear gain/loss distribution [52] and a wick-rotated dimer [53]. On the other hand, in our work, the saturation behavior of the nonlinear quadrimer system has been investigated in detail on attractor aspect. This work also addresses a crucial issue regarding choice of initial conditions while carrying out numerical simulation for such systems.

The work is organized as follows. Section 2 discusses the theoretical model and the linear eigen-spectrum of the system followed by results and discussions in Section 3. Finally, we conclude in Section 4.

## 2. Theoretical Model and linear eigen spectrum of the system

We consider a closed-form optical quadrimer waveguides system, depicted in Fig. 1. The evolution Equation describing the optical field dynamics is given by,

$$\frac{da}{dz} = -iHa \quad (1)$$

where, $a(z) = [a_1(z), a_2(z), a_3(z), a_4(z)]$ with $a_j(z)$ representing the optical field amplitude in the j-th waveguide and $H$ is the Hamiltonian of the system. Under tight-binding approximation, $H$ is given by:

$$H = \begin{pmatrix} ig & 0 & -\alpha & -\beta \\ 0 & -ig & -\gamma & -\delta \\ -\alpha & -\gamma & ig & 0 \\ -\beta & -\delta & 0 & -ig \end{pmatrix} \quad (2)$$

Here, $g$ is the gain/loss parameter, and $[\alpha, \beta, \gamma, \delta]$ are the normalized coupling constants between various waveguides as shown in Fig. 1.

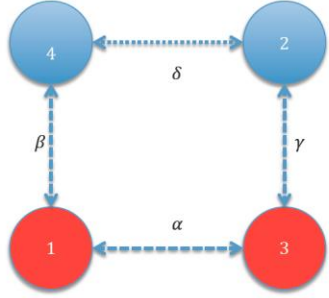

**Fig. 1.** (Color online) Schematic diagram of the system configuration under study.

The system is PT-symmetric if $[H, PT] = 0$, where $P$ is a space-reversal linear operator and $T$ performs element-wise complex conjugation [7]. The Parity operator is defined as [46]:

$$P = \begin{pmatrix} 0 & 0 & 0 & 1 \\ 0 & 0 & 1 & 0 \\ 0 & 1 & 0 & 0 \\ 1 & 0 & 0 & 0 \end{pmatrix} \quad (3)$$

We find that the system under consideration is PT-symmetric when $\alpha = \delta$. Direct diagonalization of the Hamiltonian yields the following eigenvalues:

$$\lambda_{1,2} = \pm \frac{\sqrt{f_1 + f_2}}{\sqrt{2}}, \quad \lambda_{3,4} = \pm \frac{\sqrt{f_1 - f_2}}{\sqrt{2}} \quad (4)$$

where, $f_1 = -2g^2 + \beta^2 + \gamma^2 + 2\delta^2$ and

$$f_2 = \sqrt{\beta^4 - 2\beta^2\gamma^2 + \gamma^4 - 16g^2\delta^2 + 4\beta^2\delta^2 + 8\beta\gamma\delta^2 + 4\gamma^2\delta^2}$$

It is well known that if the system is in the unbroken PT-regime, all the eigenvalues are real while if all or some of the eigenvalues are complex the system is in broken-PT symmetric regime [17]. In order to investigate the *PT* threshold of the system, in Fig. 2 we plot the appropriate eigen-spectrum of the system, taking the following coupling parameters: $\alpha = \delta = 2, \beta = 1$ and $\gamma = 4$.

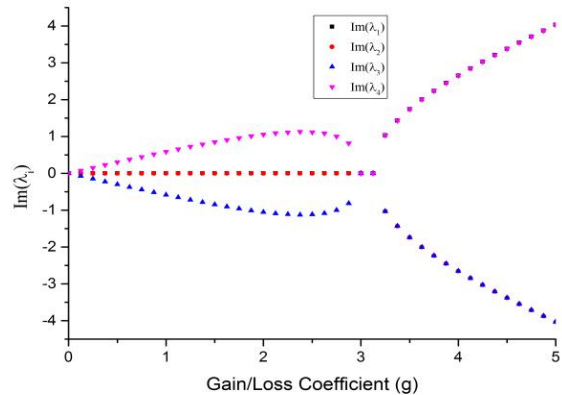

**Fig. 2.** (Color online) Eigen-spectrum of the linear system vs. loss/gain parameter

It is easy to see that we have a narrow region in $g$ where all the eigenvalues are real, i.e. a region of unbroken PT-symmetry while outside this region PT-symmetry is broken. For the chosen set of parameters, we obtain two PT-thresholds, namely $g_{th} = 3.0$ and $g_{th} = 3.125$.

On the other hand, in nonlinear regime, the propagation dynamics of the system are described by the following set of coupled nonlinear equations:

$$\frac{da_1}{dz} = ga_1 + i\alpha a_3 + i\beta a_4 + i|a_1|^2 a_1$$
$$\frac{da_2}{dz} = -ga_2 + i\gamma a_3 + i\delta a_4 + i|a_2|^2 a_2 \quad (5)$$
$$\frac{da_3}{dz} = ga_3 + i\alpha a_1 + i\gamma a_2 + i|a_3|^2 a_3$$
$$\frac{da_4}{dz} = -ga_4 + i\beta a_1 + i\delta a_2 + i|a_4|^2 a_4$$

We assume that each waveguide exhibits Kerr-nonlinearity of equal strength.

## 3. Results and Discussions

We now carry out a linear stability analysis of the nonlinear quadrimer system. The set of four coupled equations (5) can be rewritten, using the prescription, say $a_1 = x_1 + ix_2$, $a_2 = x_3 + ix_4$, and so on, as follows:

$$\dot{x}_1 = gx_1 - \alpha x_6 - \beta x_8 - (x_1^2 + x_2^2)x_2; \dot{x}_2 = gx_2 + \alpha x_5 + \beta x_7 + (x_1^2 + x_2^2)x_1$$
$$\dot{x}_3 = -gx_3 - \gamma x_6 - \alpha x_8 - (x_3^2 + x_4^2)x_4; \dot{x}_4 = -gx_4 + \gamma x_5 + \alpha x_7 + (x_3^2 + x_4^2)x_3$$
$$\dot{x}_5 = gx_5 - \alpha x_2 - \gamma x_4 - (x_5^2 + x_6^2)x_6; \dot{x}_6 = gx_6 + \alpha x_1 + \gamma x_3 + (x_5^2 + x_6^2)x_5 \quad (6)$$
$$\dot{x}_7 = -gx_7 - \beta x_2 - \alpha x_4 - (x_7^2 + x_8^2)x_8; \dot{x}_8 = -gx_8 + \beta x_1 + \alpha x_3 + (x_7^2 + x_8^2)x_7$$

This enables us to evaluate the fixed points of the system and then carry out linear stability analysis. Afterwards, we seek to find the nonlinear dynamics of the system about its fixed points. The Jacobian, $J$, of the coupled system with these new variables, after linearization, is worked out to be:

$$J = \begin{bmatrix} J_1 & J_2 \\ J_3 & J_4 \end{bmatrix} \quad (7)$$

Here, $J_1 = \begin{pmatrix} g - 2x_1 x_2 & -x_1^2 - 3x_2^2 & 0 & 0 \\ 3x_1^2 + x_2^2 & g + 2x_1 x_2 & 0 & 0 \\ 0 & 0 & -g - 2x_3 x_4 & -x_3^2 - 3x_4^2 \\ 0 & 0 & 3x_3^2 + x_4^2 & -g + 2x_3 x_4 \end{pmatrix}$, $J_2 = \begin{pmatrix} 0 & -2 & 0 & -1 \\ 2 & 0 & 1 & 0 \\ 0 & -4 & 0 & -2 \\ 4 & 0 & 2 & 0 \end{pmatrix}$, $J_3 = \begin{pmatrix} 0 & -2 & 0 & -4 \\ 2 & 0 & 4 & 0 \\ 0 & -1 & 0 & -2 \\ 1 & 0 & 2 & 0 \end{pmatrix}$ and $J_4 = \begin{pmatrix} g - 2x_5 x_6 & -x_5^2 - 3x_6^2 & 0 & 0 \\ 3x_5^2 + x_6^2 & g + 2x_5 x_6 & 0 & 0 \\ 0 & 0 & -g - 2x_7 x_8 & -x_7^2 - 3x_8^2 \\ 0 & 0 & 3x_7^2 + x_8^2 & -g + 2x_7 x_8 \end{pmatrix}$.

It is possible to obtain considerable insight about the behavior of the system from the determinant of the Jacobian and its eigenvalues evaluated at various values of the $g$-parameter. We have chosen $g = 2.5$ for the rest of analysis in this work. It is easy to see that $x_i = 0$ is the most trivial fixed point of the system. The corresponding eigenvalues of the Jacobian are (3.7081i, -3.7081i, 3.7081i, -3.7081i, 1.1180, 1.1180, -1.1180, -1.1180). In passing, it is worthwhile to note that our investigation shows that for $0 < g < 3$ and $g > 3.125$, the fixed point $x_i = 0$ is hyperbolic, as the Jacobian has complex eigenvalues. On the other hand, in the unbroken PT symmetric region, i.e. $3 < g < 3.125$, the Jacobian has purely imaginary eigenvalues, which indicates that the fixed point $x_i = 0$ is non-hyperbolic. Due to the presence of real positive eigenvalues, we can infer that this is an unstable fixed point. Clearly, choice of initial conditions in numerical simulations or power launching condition at the input ends of the waveguides, in the context of experiments now becomes a highly nontrivial issue. For example, choosing the initial condition such as: $(a_1 = 1, a_2 = a_3 = a_4 = 0)$ would result in huge perturbation in the fixed point. Using Newton Raphson method, we have obtained the following nontrivial fixed point of the system: $(-0.95224, 0.82349, 1.29818, 1.26033, 0.35102, -1.77497, -1.25158, -0.13579)$, correct up to five decimal places with error tolerance set at $10^{-10}$. The determinant of the Jacobian evaluated at the above mentioned fixed point is $\approx 1.6786 \times 10^{-7}$. From the infinitesimal value of the determinant, it can be inferred that it is not an invertible matrix. Hence, the system is not exactly linear in the vicinity of the fixed point. On the other hand, the corresponding eigenvalues of the Jacobian are purely imaginary. This implies that these fixed points are non-hyperbolic. This in turn means that only a numerical solution of the nonlinear system will give us the appropriate picture of the propagation dynamics of the optical fields along the length of the waveguide. We numerically solve Eq. (6) with perturbed fixed point as our initial conditions. For our simulation the fixed point perturbation is chosen to be: $10^{-5}$. We find that the system does have an attractor. This could be clearly seen from Fig. 3.

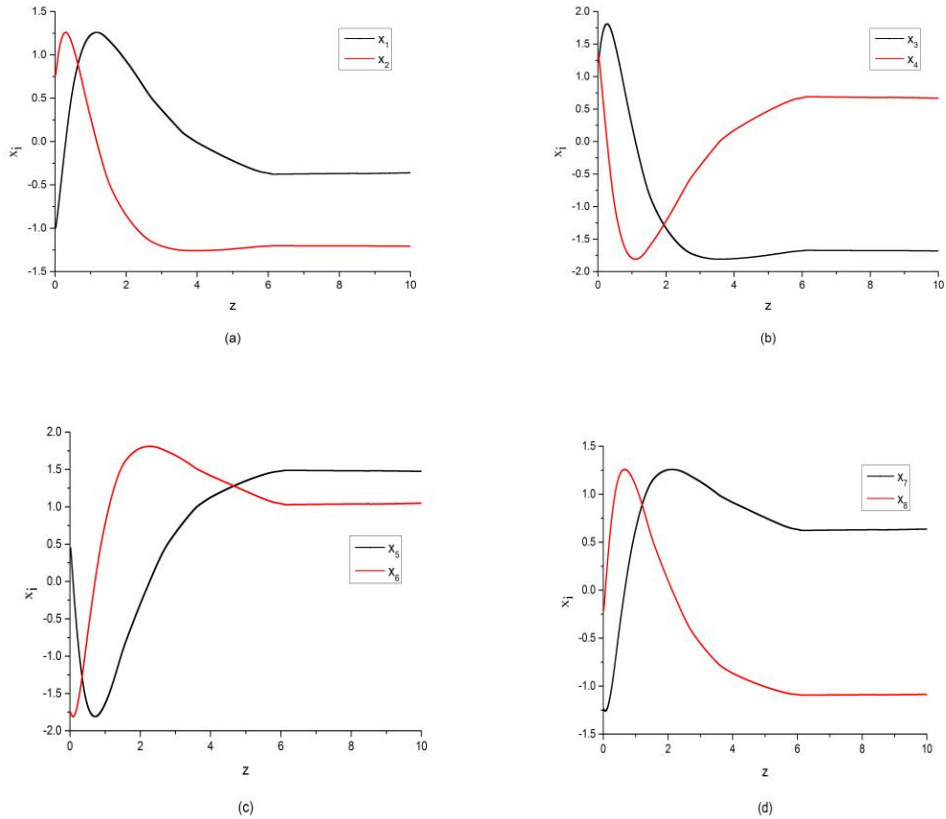

**Fig. 3.** (Color online) Spatial evolution of $x_n s$

Now it is clear that one can obtain saturation in the optical power along the propagation direction subject to judicious choice of initial conditions. For instance, in Fig. 4 we plot the spatial evolution of optical power, defined as $P_1 = |x_1 + i\, x_2|^2$, in the first waveguide.

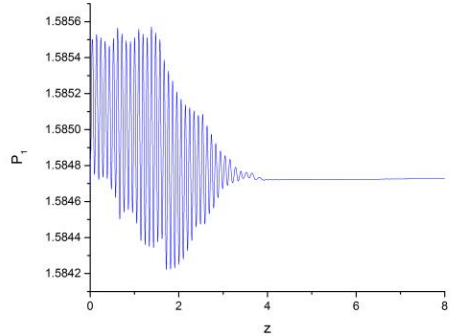

**Fig.4.** (Color Online) Spatial evolution of optical power in the first waveguide.

It could be seen from Fig. 4 that optical power gets saturated in waveguide 1. We find that similar behavior is shown by other three waveguides also. These power saturation behaviors of the system could be attributed to the attractor perspective.

## 4. Conclusions

To conclude, we have considered a closed-form parity-time symmetric quadrimer waveguides structure with a specific coupling scheme. From the linear eigen spectrum of the system it is seen that the system has a narrow window of unbroken PT-symmetric region. In nonlinear regime, for a given value of the gain/loss parameter, we have shown that the system yields stationary solutions that correspond to attractors. A detailed analysis has been provided in confirming the attractor aspect of the stationary solutions. The power saturation behaviors of the system could be attributed to the attractor perspective.

**Acknowledgments**

S. K. G. would like to thank MHRD, Government of India, for supporting through a research fellowship. J. P.D. and A.K.S. would like to acknowledge the financial support from DST, Government of India (Grant No. SB/FTP/PS-047/2013).